\documentclass[amsmath,aps,prb,reprint,twocolumn,longbibliography,superscriptaddress,floatfix]{revtex4-2}
\usepackage{graphicx}
\usepackage{amsmath}
\usepackage{amssymb}
\usepackage{bm}
\usepackage{xfrac}
\usepackage[normalem]{ulem}
\usepackage[usenames,dvipsnames]{xcolor}
\usepackage{txfonts}
\usepackage[mathscr]{euscript}
\usepackage[pdftex, pdflang={en-US},
colorlinks=true,
pdfauthor={Pontus Laurell},
pdftitle={Magnetic excitations, non-classicality and quantum wake spin dynamics in the short-range Hubbard chain}]{hyperref}

\begin{document}
	
%


\title{Magnetic excitations, non-classicality and quantum wake spin dynamics in the Hubbard chain}
\author{Pontus Laurell}
\email{plaurell@utk.edu}
\affiliation{Department of Physics and Astronomy, University of Tennessee, Knoxville, Tennessee 37996, USA.}
\affiliation{Computational Sciences and Engineering Division, Oak Ridge National Laboratory, Oak Ridge, Tennessee 37831, USA}
\author{Allen Scheie}
\affiliation{Neutron Scattering Division, Oak Ridge National Laboratory, Oak Ridge, Tennessee 37831, USA}
\author{D. Alan Tennant}
\affiliation{Department of Physics and Astronomy, University of Tennessee, Knoxville, Tennessee 37996, USA.}
\affiliation{Materials Science and Technology Division, Oak Ridge National Laboratory, Oak Ridge, Tennessee 37831, USA}
\affiliation{Quantum Science Center, Oak Ridge National Laboratory, Tennessee 37831, USA}
\affiliation{Shull-Wollan Center, Oak Ridge National Laboratory, Tennessee 37831, USA}
\author{Satoshi Okamoto}
\affiliation{Materials Science and Technology Division, Oak Ridge National Laboratory, Oak Ridge, Tennessee 37831, USA}
\affiliation{Quantum Science Center, Oak Ridge National Laboratory, Tennessee 37831, USA}
\author{Gonzalo Alvarez}
\affiliation{Computational Sciences and Engineering Division, Oak Ridge National Laboratory, Oak Ridge, Tennessee 37831, USA}
\affiliation{Center for Nanophase Materials Sciences, Oak Ridge National Laboratory, Oak Ridge, Tennessee 37831, USA}
\author{Elbio Dagotto}
\affiliation{Department of Physics and Astronomy, University of Tennessee, Knoxville, Tennessee 37996, USA.}
\affiliation{Materials Science and Technology Division, Oak Ridge National Laboratory, Oak Ridge, Tennessee 37831, USA}

\date{\today}


\begin{abstract}
	Recent work has demonstrated that quantum Fisher information (QFI), a witness of multipartite entanglement, and magnetic Van Hove correlations $G(r,t)$, a probe of local real-space real-time spin dynamics, can be successfully extracted from inelastic neutron scattering on spin systems through accurate measurements of the dynamical spin structure factor $S(k,\omega)$. Here we apply theoretically these ideas to the half-filled Hubbard chain with nearest-neighbor hopping, away from the strong-coupling limit. This model has 
	nontrivial redistribution of spectral weight in $S(k,\omega)$ going from the non-interacting limit ($U=0$) to  
	strong coupling 
	($U\rightarrow \infty$), where it reduces to the Heisenberg quantum spin chain. 
	We use the density matrix renormalization group (DMRG) to find $S(k,\omega)$, from which QFI is then calculated. 
	We find that QFI grows with $U$. With realistic energy resolution it becomes capable of witnessing bipartite entanglement above $U=2.5$ (in units of the hopping), where it also changes slope. 
	This point is also proximate to 
	slope changes of the bandwidth $W(U)$ and the half-chain von Neumann entanglement entropy. We compute $G(r,t)$ by Fourier-transforming $S(k,\omega)$. 
	The results indicate a crossover in the short-time short-distance dynamics at low $U$ characterized by ferromagnetic lightcone wavefronts, to a Heisenberg-like behavior at large $U$ featuring antiferromagnetic lightcones and spatially period-doubled antiferromagnetism. We find this crossover has largely been completed by $U=3$. Our results thus provide evidence that, in several aspects, the strong-coupling limit of the Hubbard chain is reached qualitatively already at a relatively modest interaction strength.  
	We discuss experimental candidates for observing the $G(r,t)$ dynamics found at low $U$.
\end{abstract}
\maketitle

\section{Introduction}
In principle, the full information about a quantum many-body system at a given time $t$ is contained in the set of all equal-time correlation functions \cite{PhysRevX.10.011020}. However, most current experimental scattering and spectroscopy methods only access one- and two-point correlation functions. It is thus of interest to extract as much information as we can from these more accessible correlators, especially measures of quantum correlations or ``quantumness'' in strongly correlated systems \cite{Chiara2018}. Indeed, accessing more information on the quantum states from measurements would aid in identifying and selecting materials and models for further study as well as help in the design of more effective experiments. 

For example, inelastic neutron scattering probes magnetic excitations by measuring spin-spin correlations encoded in the dynamical spin structure factor $S(\mathbf{k},\omega)$ \cite{Squires, Lovesey1986}. Recently, it was shown that a witness \cite{RevModPhys.80.517, Vedral2008, Guehne2009} of multipartite entanglement known as quantum Fisher information (QFI) \cite{PhysRevA.85.022321, PhysRevA.85.022322}  can be obtained from an integral over $S(\mathbf{k},\omega)$ \cite{Hauke2016}. 
Similarly to how Bell inequalities have been used to demonstrate entanglement in few-particle systems \cite{PhysicsPhysiqueFizika.1.195, Aspect_1999, RevModPhys.71.S288, RevModPhys.81.865} and certain many-particle systems \cite{Schmied16, PhysRevLett.118.140401}, one can obtain inequalities for QFI and other entanglement witnesses that can only be satisfied in specific classes of entangled states. Witnesses thus provide a promising approach to entanglement detection and quantification. 
Neutron measurements of QFI temperature scaling \cite{PhysRevResearch.2.043329, PhysRevB.103.224434} and QFI entanglement bounds \cite{PhysRevB.103.224434, PhysRevLett.127.037201, Scheie2021} have since been made for low-dimensional quantum spin systems. Other entanglement witnesses relying on other subsets of the information in $S(\mathbf{k},\omega)$, such as the static structure factor \cite{PhysRevLett.103.100502, PhysRevLett.106.020401, Frerot2021} or equal-time real-space spin-spin correlations \cite{PhysRevA.61.052306, PhysRevA.69.022304, PhysRevLett.93.167203}, have also been discussed and in some cases measured with neutrons \cite{PhysRevB.103.224434, PhysRevLett.127.037201, Scheie2021, PhysRevA.73.012110, PhysRevLett.99.087204, Garlatti2017}. The use of such witnesses for, {\it e.g.}, helping to experimentally identify quantum spin liquid candidates is actively being considered \cite{Scheie2021}. 

An alternative perspective on $S(\mathbf{k},\omega)$ may be found by recalling that it is a Fourier transform of an underlying real-space two-site two-time correlation function $G(\mathbf{r},t)$. In neutron scattering $G(\mathbf{r},t)$ is known as the Van Hove correlation function \cite{Squires, PhysRev.95.249, PhysRev.95.1374}, whereas in the context of lattice models it is more commonly called a dynamical correlation function [for example, in a spin-isotropic one-dimensional system we define $G(r,t)\sim \langle S_i^z(0)S_{i+r}^z(t)\rangle$].  
Investigating this quantity instead of the momentum- and frequency-resolved dynamics usually studied in spectroscopy might lead to new insights, perhaps particularly in correlated systems with local interactions.

As noted by Van Hove in 1954 \cite{PhysRev.95.249, PhysRev.95.1374} the imaginary part, $\mathrm{Im}\left[ G(\mathbf{r},t) \right],$ vanishes for a classical system. 
Non-zero $\mathrm{Im}\left[ G(\mathbf{r},t) \right]$ thus indicates quantum properties of the local dynamics. The 
case $\mathrm{Im}\left[ G(\mathbf{r},t) \right]\equiv 0$ is relevant to, e.g., classical fluids, where $G(\mathbf{r},t)$ may be understood as the average number density at $\mathbf{r}$ and $t$ given that a particle was at the origin at time $t^\prime=0$. This picture has been used to interpret experimental data on liquid lead \cite{PhysRevLett.3.259}, water \cite{Iwashitae1603079} and Zr$_{80}$Pt$_{20}$ \cite{doi:10.1063/1.5144256, doi:10.1063/5.0024013}. Here we are instead concerned with the quantum case, where the interpretation of $G(\mathbf{r},t)$ is complicated by the noncommutativity of operators at different times. This may explain, in part, why numerical results for $G(\mathbf{r},t)$ are rarely reported in the literature---see Ref.~\cite{PhysRevB.94.085136} for an exception to this rule---despite such real-time correlators being used as intermediate steps in some numerical techniques to calculate $S(\mathbf{k},\omega)$. Recently such correlations have also been obtained using quantum computers \cite{Chiesa2019, PhysRevB.101.014411, PRXQuantum.2.010317}.

Importantly, this noncommutativity allows a window into the dynamics of local quasiparticles and the quantum coherence of the system. This was recently reported  
in the $S=1/2$ Heisenberg antiferromagnetic chain system KCuF$_3$, where $G(r,t)$ was obtained from both theory and neutron scattering \cite{Scheie2022}. The data showed the propagation of correlations was limited by a lightcone, as expected from Lieb-Robinson bounds \cite{Lieb1972, 10.21468/SciPostPhys.5.5.052, PRXQuantum.1.010303}. In addition, a signal oscillating in time with fixed period was found, and $\mathrm{Re}[G(r,t)]$ revealed dynamic spatial period-doubling in the antiferromagnetic correlations, which could heuristically be interpreted in terms of spinon-spinon scattering processes. This dynamic period-doubled state above the lightcone is markedly different from the static N\'eel order below it, featuring vanishing odd-nearest neighbor correlations, and was referred to as a ``quantum wake'' in analogy with the wake created by a moving ship. 
Finally, $G(r,t)$ was found to tend towards real non-negative values with temperature, indicating a crossover to classical behavior.

In the current work we apply these QFI and $G(r,t)$ approaches to one of the simplest models of strongly correlated electrons: the paradigmatic 1D Hubbard chain with repulsive on-site interactions and nearest-neighbor hopping \cite{PhysRevLett.10.159, Hubbard1963, Kanamori1963, essler2005one}. It interpolates between a noninteracting system with a planewave solution at weak coupling and the Heisenberg spin chain at strong coupling \cite{Heisenberg1928, Auerbach1994}. We use the density matrix renormalization group (DMRG) \cite{PhysRevLett.69.2863, PhysRevB.48.10345} to obtain $S(k,\omega)$, QFI, and $G(r,t)$ as a function of the Hubbard interaction strength.
We find crossovers in spectra, $G(r,t)$, QFI, and entanglement entropy. These all occur at modest Hubbard interaction strengths, $U<W(0)$, where $W(0)=4\tilde{t}$ is the non-interacting bandwidth, reflecting a crossover from itinerant electron physics to increasingly localized magnetic moments as $U$ is raised.

This paper is organized as follows. In Sec.~\ref{sec:quantities} we fix our notation for $S(\mathbf{k},\omega)$, and introduce the definitions of QFI and $G(\mathbf{r},t)$ used in this work. In Sec.~\ref{sec:model} we review relevant results for the 1D Hubbard model and discuss the numerical method. Our results are presented in Sec.~\ref{sec:results} and discussed in Sec.~\ref{sec:discussion} along with potential material realizations and experimental considerations. We summarize the conclusions in Sec.~\ref{sec:conclusion}. Some technical details and analytical results are contained in appendices.

\section{Quantum Fisher information and Van Hove correlations}\label{sec:quantities}

The dynamical spin structure factor (DSF) is defined by 
\begin{align}
	&S^{ab}	\left( \mathbf{k}, {\hbar}\omega\right)	=	\frac{1}{2\pi N\hbar} \int_{-\infty}^\infty \sum_{i,j} \left\langle S_i^a \left(t\right) S_j^b (0) \right\rangle e^{-i\mathbf{k} \cdot \left( \mathbf{r}_i - \mathbf{r}_j \right)} e^{+i\omega t} \mathrm{d}t\nonumber\\
												&= \frac{1}{2\pi N\hbar} \int_{-\infty}^\infty \sum_{i,j} \left\langle S_i^a \left(0\right) S_j^b (t) \right\rangle e^{-i\mathbf{k} \cdot \left( \mathbf{r}_i - \mathbf{r}_j \right)} e^{-i\omega t} \mathrm{d}t,	\label{eq:sqw}
\end{align}
where $N$ is the number of sites $i,j$. $a,b\in\{x,y,z\}$, $\hbar=h/2\pi$ is the reduced Planck constant, and, in the Heisenberg picture, $S_j^b(t) = e^{iHt/\hbar}S_j^b e^{ -iHt/\hbar}$. Both conventions shown in Eq.~\eqref{eq:sqw} are used in the literature and are equivalent assuming stationarity holds, i.e. $\left\langle S_i^a \left(t\right) S_j^b (0) \right\rangle = \left\langle S_i^a \left(0\right) S_j^b (-t) \right\rangle$. 
The DSF satisfies detailed balance, $S(\mathbf{k},-\hbar\omega)=\exp \left( -\hbar\omega/k_B T\right) S(\mathbf{k},\hbar\omega)$, and is constrained by sum rules, such as
\begin{align}
	\sum_a	\int_{-\infty}^\infty \mathrm{d}\left( \hbar\omega \right) \int_\mathrm{B.Z.} \mathrm{d}\mathbf{k} S^{aa}\left( \mathbf{k}, \hbar\omega \right)	\frac{V_0}{(2\pi)^d} &=	S(S+1),	\label{eq:sumrule}
\end{align}
where $\mathrm{B.Z.}$ indicates the integral is taken over the first Brillouin zone, {$a\in \{ x,y,z\}$}, $d$ is number of spatial dimensions, and $V_0$ is the volume of the unit cell of the direct lattice. 
[${V_0}/{(2\pi)^d}=1$ if momenta are measured in lattice units $h,k,l,\dots$.]  
Equation~\eqref{eq:sumrule} applies to systems with spin-$S$ moments on each site. This assumption holds for the Hubbard model at half-filling and strong coupling, but breaks down at finite electron-electron repulsion, for which 
sites may be unoccupied or doubly occupied. The corrected sum rule for one-band electronic systems is \cite{PhysRevB.72.224511}
\begin{align}
	\sum_a	\int_{-\infty}^\infty \mathrm{d}\left( \hbar\omega \right) \int_\mathrm{B.Z.} \mathrm{d}\mathbf{k} S^{aa}\left( \mathbf{k}, \hbar\omega \right)	\frac{V_0}{(2\pi)^d}&=	\frac{3}{4} \left( n-2D\right),	\label{eq:sumrule:hubbard}
\end{align}
where
\begin{align}
	n	&=	\frac{1}{N} \sum_{i,\sigma} \langle n_{i\sigma}\rangle,\\
	D	&=	\frac{1}{N} \sum_i \langle n_{i\uparrow} n_{i\downarrow}\rangle,    \label{eq:doubleoccupation}
\end{align}
measure the average orbital and double occupancy, respectively. (We have assumed the electron $g$-factor $g_e\approx 2$, and used units where $\mu_B=1$. To restore these factors explicitly, see full expressions in Ref.~\cite{PhysRevB.72.224511}.) The DSF is related to the dynamical susceptibility $\chi\prime\prime$ 
by the fluctuation-dissipation theorem $\chi\prime\prime\left(\mathbf{k},\hbar\omega,T\right)=\tanh\left(\sfrac{\hbar\omega}{2k_BT}\right)\tilde{S}\left(\mathbf{k},\hbar\omega\right)$, where $\tilde{S}\left( \mathbf{k},\hbar\omega\right) = S\left( \mathbf{k},\hbar\omega\right) + S\left( \mathbf{k},-\hbar\omega\right)$
\footnote{The fluctuation-dissipation relation states $S\left(\mathbf{k},\hbar\omega\right)=\left( 1-e^{-\hbar\omega/k_BT}\right)^{-1} \chi\prime\prime\left(\mathbf{k},\hbar\omega,T\right)$. Detailed balance yields $S\left(\mathbf{k},-\hbar\omega\right)=\left( e^{\hbar\omega/k_BT}-1\right)^{-1} \chi\prime\prime\left(\mathbf{k},\hbar\omega,T\right)$ and thus $S\left(\mathbf{k},\hbar\omega\right)+S\left(\mathbf{k},-\hbar\omega\right)=\coth \left( \frac{\hbar\omega}{2k_BT}\right) \chi\prime\prime\left(\mathbf{k},\hbar\omega,T\right)$.}.

The QFI density may be written as follows: \cite{Hauke2016, PhysRevB.103.224434, PhysRevLett.127.037201}
\begin{equation}
f_\mathcal{Q}({\mathbf{k},T}) = \frac{4}{\pi} \int_0^{\infty} \mathrm{d} (\hbar \omega) \tanh \left( \frac{\hbar \omega}{2 k_B T} \right) \chi\prime\prime (\mathbf{k}, \hbar \omega, T) \label{eq:QFI}.
\end{equation}
Note that quantitative determination of QFI requires working with absolute intensities, which is ensured by proper normalization of $S(\mathbf{k},\hbar\omega)$ according to appropriate sum rules. We will use the sum rule \eqref{eq:sumrule:hubbard}. Following Ref.~\cite{PhysRevB.103.224434} we introduce the normalized QFI (nQFI),
\begin{equation}
	\mathrm{nQFI} (\mathbf{k},T)	= \frac{f_\mathcal{Q}(\mathbf{k},T)}{12S^2},	\label{eq:normalizedQFI}
\end{equation}
where the spin length $S=1/2$ for electrons. These quantities become useful because (i) they are experimentally accessible, and (ii) it is possible to derive bounds for $f_\mathcal{Q}$ (or nQFI) that can only be reached by certain classes of multipartite-entangled states \cite{PhysRevA.85.022321, PhysRevA.85.022322, Pezze2014}. 
The bound applicable to unpolarized inelastic neutron scattering on magnetic systems indicates that we witness \emph{at least} $(m+1)$-partite entanglement when $\mathrm{nQFI}>m$, where $m$ is an integer and divisor of the system size \cite{PhysRevLett.127.037201}. The integer $m$ is known as the entanglement depth, and represents the minimum number of entangled sites. 
The existence of such bounds can intuitively be understood by noting that entangled states can have stronger spin-spin correlations at fixed momentum $\mathbf{k}$ than any separable (i.e. non-entangled) state, which translates to high spectral weight integrated over in Eq.~\eqref{eq:QFI}. Similarly, the more sites are entangled, stronger and stronger spin-spin correlations become possible. 

The derivation of the bound makes no assumption about the nature of the system for which $S(\mathbf{k},\hbar\omega)$ is observed or calculated, it relies only on $S(\mathbf{k},\hbar\omega)$ being a dynamical correlation associated with local and bounded Hermitian operators (here, spin operators) \cite{PhysRevA.85.022321, PhysRevA.85.022322, Hauke2016, Pezze2014}. In the Hubbard chain, the spin operators are implemented through
\begin{align}
	S_i^a	&=	\frac{\hbar}{2} c_{i\sigma}^\dagger \sigma^a_{\sigma\sigma'} c_{i\sigma'},
\end{align}
where $\vec{\sigma}=(\sigma_x,\sigma_y,\sigma_z)$ is the vector of Pauli matrices and $c_{i,\sigma}^\dagger$ creates an electron of spin $\sigma\in \{\uparrow,\downarrow\}$ at site $i$, resulting in the same bound as for local moment spin-$1/2$ systems. Alternatively, this follows because the only allowed spin states are $0$ (for empty or double-occupied sites) and $1/2$ (for singly occupied sites).  
We note that these assumptions do not hold for all response functions. For example, the spectral function $A(\mathbf{k},\hbar\omega)$ is a dynamical correlation associated with the \emph{non}-Hermitian operators $c,c^\dagger$. On the other hand, e.g. the dynamical charge structure factor $N(\mathbf{k},\hbar\omega)$ \cite{PhysRevB.85.165132}, associated with the density operator $n$, does meet the assumptions.

Note also that QFI defined this way will probe entanglement carried by spin correlations. Electronic systems also have entanglement in the charge sector, which we expect to dominate at low $U$,  due to the Pauli exclusion principle. Other probes would be needed to access this type of entanglement. 
In the Hubbard chain, we focus on the entanglement associated with the antiferromagnetic wave vector $k=\pi$, since staggered magnetization is a relevant operator in the renormalization group sense \cite{Hauke2016, PhysRevLett.127.037201}.

By definition Eq. \eqref{eq:sqw}, $S(\mathbf{k},\hbar\omega)$ is a Fourier transform of a two-point two-time correlation function. Assuming the system is translation invariant we have
\begin{align}
	G^{ab} (\mathbf{r}=\mathbf{r}_i-\mathbf{r}_j,t)	&=	\left\langle S_i^a \left(0\right) S_j^b (t) \right\rangle,
\end{align}
which we will refer to as a Van Hove correlation in analogy with terminology used in the context of neutron scattering on liquids \footnote{It is also known as a pair-correlation function \cite{Lovesey1986}. We avoid this name since it may be confused for the (pair)-pair correlation functions used to describe superconductivity.}. Since $S(\mathbf{k},\hbar\omega)$ is real-valued, $G(\mathbf{r},t)=G^\star (-\mathbf{r},-t)$.  The real (imaginary) part may be written with an anticommutator (commutator),
\begin{align}
	\mathrm{Re} \left[ G^{ab}(\mathbf{r},t) \right]	&=	\frac{1}{2} \left\langle \left\{ S_i^a(0), S_j^{b}(t)\right\} \right\rangle,\\
	\mathrm{Im} \left[ G^{ab}(\mathbf{r},t) \right]	&=	\frac{1}{2i} \left\langle \left[ S_i^a(0), S_j^{b}(t)\right] \right\rangle,
\end{align}
implying that the imaginary part (i) vanishes in a classical system (where operators are replaced by $c$-numbers), and (ii) is directly related to the dissipative susceptibility. The values of $\mathrm{Re} \left[ G^{ab}(\mathbf{r},t) \right]$ and $\mathrm{Im} \left[ G^{ab}(\mathbf{r},t) \right]$ may be related through detailed balance or fluctuation-dissipation theorems \cite{PhysRevLett.4.239, Egelstaff1965}. Additional properties of $G(\mathbf{r},t)$ are stated in Appendix~\ref{app:properties}. 

For systems described by local Hamiltonians, the imaginary part is generically expected to satisfy a Lieb-Robinson bound \cite{Lieb1972, 10.21468/SciPostPhys.5.5.052, PRXQuantum.1.010303}, such that it decays exponentially outside a light-cone determined by a system-dependent Lieb-Robinson velocity. The bound thus prevents superluminal information propagation \cite{PhysRevLett.97.050401}. 
The real part, being an anticommutator, is better viewed as a statistical property. It is expected to depend on the initial state and can have richer behavior near and outside the light-cone \cite{10.21468/SciPostPhys.5.5.052, PhysRevB.88.094306}. Since it is a statistical property, nonzero correlations outside the light-cone do not imply noncausality.

\section{Model and methods}\label{sec:model}
\begin{figure*}
	\includegraphics[width=\textwidth]{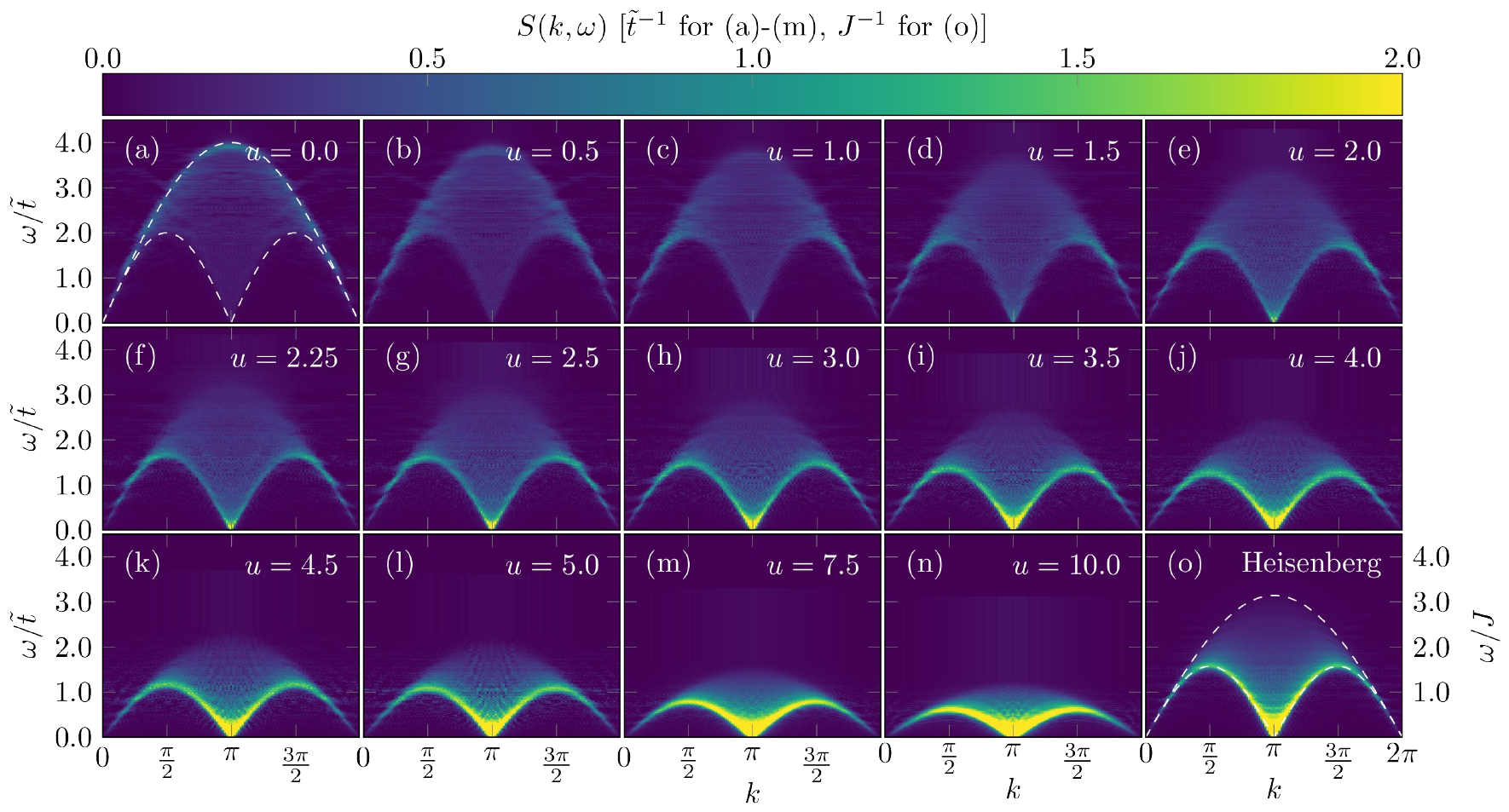}
	\caption{\label{fig:skw}$S(k,\omega)$ for the Hubbard chain as function of $u=U/\tilde{t}$ and the Heisenberg chain for chains of length $L=128$ with broadening parameter $\eta=0.05\tilde{t}$. As $u$ increases, the bandwidth shrinks and spectral weight moves from the top of the dispersion towards its bottom edge. The resulting spectrum approaches that of the Heisenberg chain shown in panel (o).
	}
\end{figure*}
\begin{figure*}
	\includegraphics[width=\textwidth]{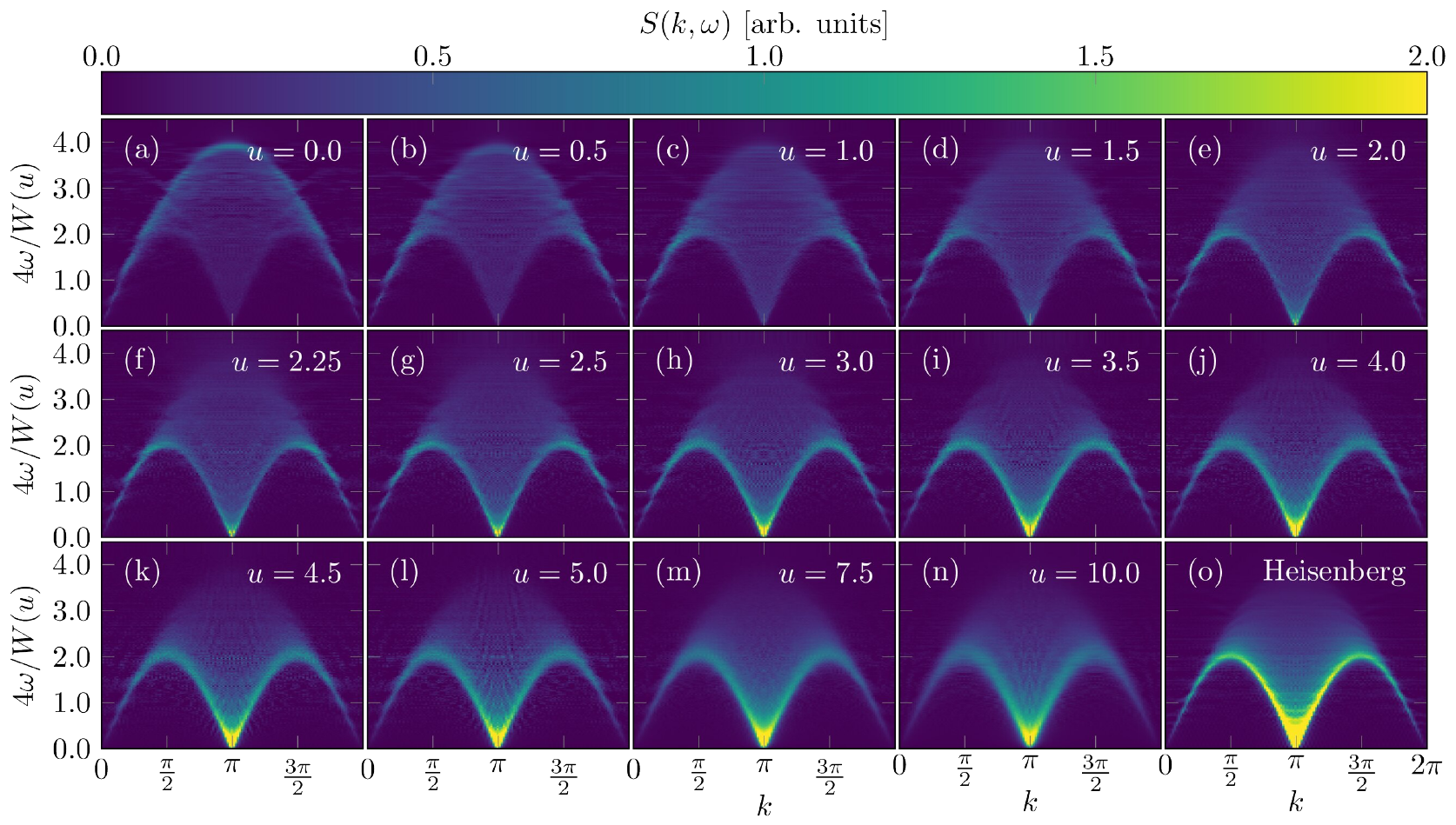}
	\caption{\label{fig:skw:scaled}$S(k,\omega)$ for the Hubbard chain as function of $u=U/\tilde{t}$ and the Heisenberg chain for chains of length $L=128$ with broadening parameter $\eta=0.05\tilde{t}$. Note that the panels have been plotted with different $u$-dependent energy scale factors (see Table~\ref{table:hubbardbandwidth}) in order to keep the apparent bandwidth constant. The redistribution of spectral weight from the top of the scattering continuum to the bottom is apparent already at relatively low values of $u$.
	}
\end{figure*}
The Hubbard model is written \cite{PhysRevLett.10.159, Hubbard1963, Kanamori1963, essler2005one}
\begin{align}
H	&=	-\tilde{t} \sum_{j=0}^{L-1}\sum_{\sigma} \left[ c_{j,\sigma}^\dagger c_{j+1,\sigma}^{} + \mathrm{H.c.} \right] + U \sum_{j=0}^{L-1} n_{j,\uparrow} n_{j,\downarrow}, \label{eq:Hubbard}
\end{align}
where $\tilde{t}$ represents the electronic nearest-neighbor hopping strength \footnote{The usual notation where $t$ denotes hopping is avoided, since $t$ will be used to denote the time variable.}, $U\geq 0$ is the on-site Hubbard repulsion, 
and $\mathrm{H.c.}$ denotes Hermitian conjugate. Throughout 
this work we will assume half-filling, use the dimensionless interaction strength $u=U/\tilde{t}$, and often take $\tilde{t}$ as our energy unit. In the following we will also work in units where $\hbar=1$. At $u=0$ the model describes non-interacting electrons, and as $u\rightarrow \infty$ the model reduces to the antiferromagnetic Heisenberg chain with only spin degrees of freedom \cite{Heisenberg1928, Auerbach1994}. The latter model was famously solved in one dimension by Bethe \cite{Bethe1931}. Utilizing that Eq.~\eqref{eq:Hubbard} conserves the number of electrons $N$ and number of down spins $M$, Lieb and Wu \cite{PhysRevLett.20.1445, PhysRevLett.21.192.2} later solved the Hubbard chain using a nested Bethe ansatz \cite{PhysRevLett.19.1312}.

Although many ground state properties can be obtained exactly, the dynamical spin correlations $S(k,\omega)$ remain a challenge. They were obtained from the exact solution to high accuracy for the Heisenberg chain \cite{PhysRevB.55.12510, Caux_2006, Mourigal2013, PhysRevLett.111.137205}, but for the Hubbard chain at finite $u$ only partial analytical results exist 
\cite{PhysRevB.59.1734}. Based on a perturbative approach, Bhaseen et al. \cite{PhysRevB.71.020405} found that there is a downward shift of spectral intensity as $u$ is increased, such that the main concentration of spectral weight moves from the top of the spectrum towards the bottom. The same intensity redistribution is seen in DMRG calculations 
\cite{PhysRevB.75.205128, PhysRevB.94.205145}, which also reveal that the itineracy effects rapidly diminish. In fact, the 
spectrum at $u=3$ is already close to that of the Heisenberg chain. Qualitatively similar results were recently obtained using a cluster perturbation theory \cite{PhysRevB.101.075122}.

We use the DMRG \cite{PhysRevLett.69.2863, PhysRevB.48.10345} as implemented in the DMRG++ software \cite{Alvarez2009}, working at zero temperature. We work with even-length chains and in the zero magnetization sector in accord with Lieb's theorem \cite{PhysRevLett.62.1201}. We calculate $S(k,\omega)$ in the Krylov correction-vector approach \cite{PhysRevB.60.335, PhysRevB.66.045114, PhysRevE.94.053308},
which works directly in frequency space. Formally this is achieved by evaluating the average in Eq.~\eqref{eq:sqw} in the ground state, employing the Heisenberg picture, and introducing an infinitesimal Lorentzian broadening $\eta>0$ to regularize the time integral,
\begin{align}
&S^{ab}(k,\omega)	=	\frac{1}{2\pi N} \int_{-\infty}^\infty \sum_{i,j} \langle \psi_0 | S_i^a e^{-i\left( H - \omega - \epsilon_0 \right) t - \eta|t|} S_j^b | \psi_0 \rangle\nonumber\\
& \quad\quad\quad\quad\quad e^{-ik (r_i-r_j)}  dt \nonumber\\
&=	\frac{i}{2\pi N} \sum_{i,j} e^{-ik (r_i-r_j)} \left( \langle \psi_0 | S_i^a \left[ H - \omega - \epsilon_0 + i \eta\right]^{-1} S_j^b | \psi_0 \rangle \right. \nonumber\\
&-\left. \langle \psi_0 | S_i^a \left[ H - \omega - \epsilon_0 - i \eta\right]^{-1} S_j^b | \psi_0 \rangle \right),	\label{eq:general}
\end{align}
where $\epsilon_0=\left\langle \psi_0 \middle| H \middle| \psi_0 \right\rangle$ is the ground state energy. 
In the diagonal case $a=b$ this expression simplifies to
\begin{align}
	S^{aa}(k,\omega)	&= -\frac{1}{\pi N} \sum_{i,j} e^{-ik (r_i-r_j)} \nonumber\\
						&\mathrm{Im} \bigg[  \langle \psi_0 | S_i^a \left[ H - \omega - \epsilon_0 + i \eta\right]^{-1} S_j^a | \psi_0 \rangle \bigg],	\label{eq:skomega:diagonal}
\end{align}
where the repeated index $a$ is not summed over. Due to spin SU(2) symmetry of the Hubbard model it is sufficient to compute only $S^{zz}(k,\omega)$.

In the numerical calculation we use finite-size chains and employ the center-site approximation, in which the sum over sites $j$ is restricted to a center site $c=L/2$. This approximation reduces the computational cost by an order of $L$ and is exact in the thermodynamic limit, but can introduce ``ringing'' artifacts in finite systems. We thus need to evaluate
\begin{align}
	S^{aa}_{j,c}(\omega) = -\frac{1}{\pi}\mathrm{Im} \bigg[  \langle \psi_0 | S_j^a \left[ H - \omega - \epsilon_0 + i \eta\right]^{-1} S_c^a | \psi_0 \rangle \bigg],
\end{align}
for each site $j$, which is achieved as described in Ref.~\cite{PhysRevE.94.053308}. Finally, since we sum over only one site index, Eq.~\eqref{eq:skomega:diagonal} is modified to read
\begin{align}
	S^{aa}(k,\omega)	&=	\frac{1}{\sqrt{L}} \sum_{j=0}^{L-1}	\cos \left[ k \left( r_i-r_c\right)\right] S^{aa}_{j,c}(\omega).
\end{align}
The cosine is appropriate for periodic boundary conditions, or open chains that are symmetric around the center site. Here we use the cosine also for chains of even length with open boundary conditions and a single center site, which introduces a small error that vanishes in the thermodynamic limit.

In the numerical computations $\eta$ is finite, and represents the half width at half maximum (HWHM) of the Lorentzian energy broadening. Its optimal value is limited by finite size according to $\eta \propto 1/L$ \cite{PhysRevB.66.045114, PhysRevE.94.053308}. In systems with gapless excitations a finite $\eta$ may introduce spurious inelastic ($\omega>0$) intensity due to an elastic peak at $\omega=0$ . To avoid this we isolate the purely inelastic intensity by subtracting from the normalized $S(k,\omega)$ a Lorentzian of broadening $\eta$ and height $S(k,0)$ at each $k$-point. The subtraction is done after normalization to the sum rule Eq.~\eqref{eq:sumrule:hubbard}. The resulting inelastic scattering is used to determine the QFI. $G(r,t)$ is calculated by an inverse Fourier transform, as in Ref.~\cite{Scheie2022}. Again, because of the SU(2) symmetry it is sufficient to consider the longitudinal part $G^{zz}(r,t)	=	\langle S_i^z (0) S_{i+r}^z (t) \rangle$.

Our results for $L=128$ sites were obtained keeping up to $m=1600$ DMRG states, achieving truncation errors below $10^{-8}$. Explicit reorthogonalization was used for all Lanczos steps in the ground state runs. For the dynamics runs, we used $\eta=0.05\tilde{t}$ and scaled the frequency step $\Delta \omega$ such that the number of sampled frequencies within the bandwidth predicted by the Bethe ansatz (see Appendix~\ref{app:bandwidth}) was kept constant and equal to $160$, with additional frequencies sampled above the predicted bandwidth. For other system sizes we scaled $m \propto L$ and $\eta \propto 1/L$. For $u\geq 7.5$ ( $u\leq 7.5$) a total of $300$ ($200$) Krylov steps were used. The increased number of steps at high $u$ was found necessary to avoid artifacts in the continuum scattering.

\section{Results}\label{sec:results}
Below we present results for the dynamical spin structure factor, as well as the quantum Fisher information and real-space real-time Van Hove correlations obtained from said DSF.

\subsection{Dynamical spin structure factor}\label{sec:results:DSF}
In Fig.~\ref{fig:skw} we show the calculated $S(k,\omega)$ as a function of $u$. 
The dashed white lines at $u=0$ in panel (a) enclose the non-interacting bandwidth between upper and lower boundaries
\begin{align}
\omega_u^{u=0}(k)	&=	4\tilde{t} \left| \sin(k/2)\right|,\\
\omega_l^{u=0}(k)	&=	2\tilde{t} \left| \sin(k)\right|.
\end{align}
Dashed white lines in Fig.~\ref{fig:skw}(o) represent the upper and lower boundaries of the two-spinon continuum for the isotropic Heisenberg antiferromagnetic chain \cite{10.1143/PTP.41.880, doi:10.1063/1.30115, Muller1981},
\begin{align}
\omega_u^\mathrm{Heisenberg}	&=	\pi J \sin\left( \frac{k}{2}\right),\\
\omega_l^\mathrm{Heisenberg}	&=	\frac{\pi J}{2} \sin (k).
\end{align}
We find the spectra follow the trend observed in Refs.~\cite{PhysRevB.71.020405, PhysRevB.75.205128, PhysRevB.94.205145}. That is, at $u=0$ the spectral weight is concentrated at the top of the spectrum. As $u$ is increased, spectral weight gets redistributed towards the bottom of the spectrum, which at strong coupling corresponds to the des-Cloizeux-Pearson dispersion \cite{PhysRev.128.2131} for a Heisenberg antiferromagnetic chain with exchange strength $J$. Second order perturbation theory in the strong coupling limit predicts 
$J=4\tilde{t}/u$. 
If $\tilde{t}$ is treated as a constant energy scale, it follows that the bandwidth quickly diminishes with $u$. It is possible to find the bandwidth of $S(k,\omega)$ at all $u$ from the Bethe ansatz (see Appendix~\ref{app:bandwidth}). If we scale all energy scales ($u$, $\eta/\tilde{t}$) such that the the bandwidth $W(u)$ is kept equal to the non-interacting bandwidth [$W(u=0)=4|\tilde{t}|=4$] we obtain the spectra in Figure \ref{fig:skw:scaled}, from which the redistribution of spectral weight is easier to see.

Due to a finite-size effect the reliability of calculated spectra decreases at large $u$, so here we report $S(k,\omega)$ for $u \leq 10$. The finite-size effect may be understood as follows. As $u$ increases, the bandwidth becomes small and eventually comparable to the broadening $\eta$. As previously mentioned, the optimal broadening is limited by system size. Thus, for fixed $L$, the ratio $W(u)/\eta$ becomes too small at large $u$ to allow reliable QFI results (as QFI is an integral over $S(k,\omega)$ washing-out of the spectrum becomes an issue). Our $L=128$ results were obtained with $\eta=0.05\tilde{t}$, for which $W(10)/\eta\approx 24$ and $W(15)/\eta\approx 16$. There is no obvious value to choose as cutoff for $W(u)/\eta$, but here we require $W(u)/\eta>20$.

\subsection{Quantum Fisher information}
\begin{figure}
	\includegraphics[width=\columnwidth]{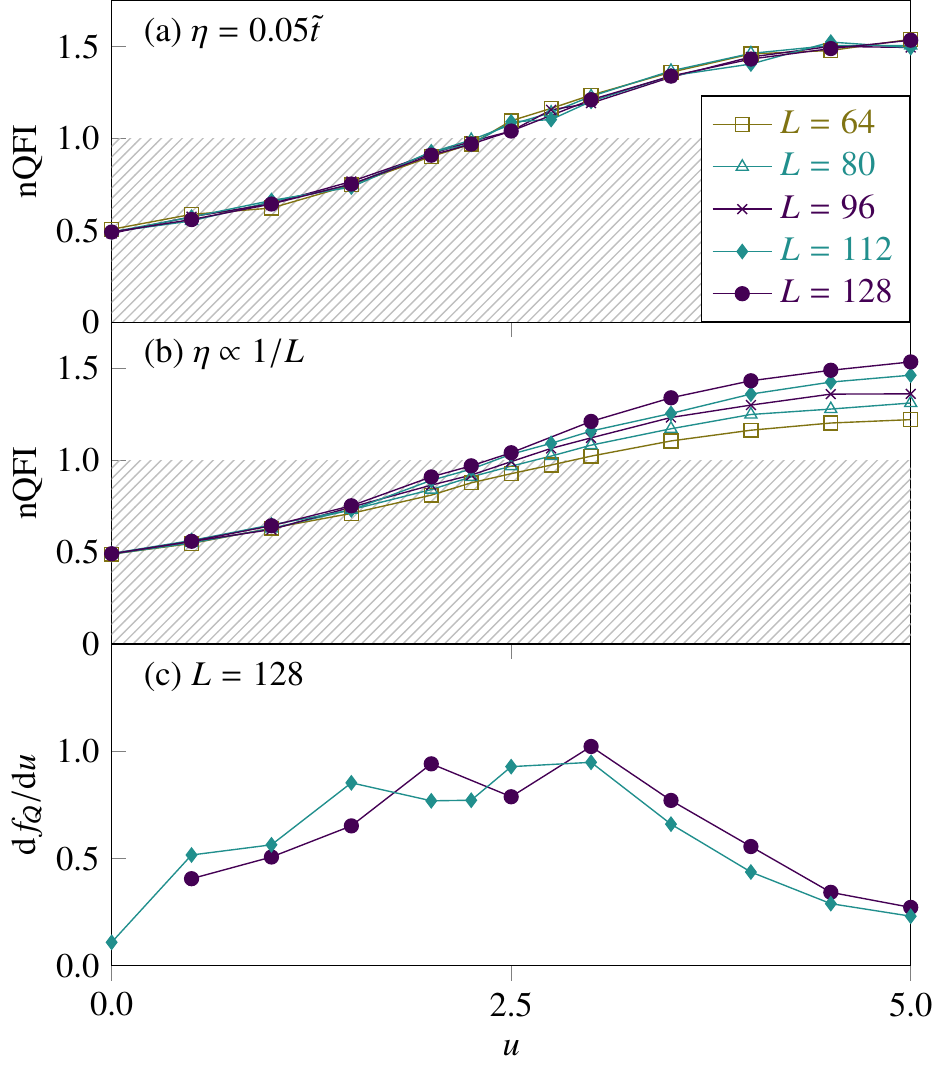}
	\caption{\label{fig:QFI}(a) The normalized QFI [nQFI] as function of $u=U/\tilde{t}$, calculated from $S(k=\pi,\omega)$ where elastic contributions have been removed and with $\eta=0.05\tilde{t}$ for several system sizes. For nQFI$>1$ (i.e. outside the shaded region), at least bipartite entanglement is witnessed by QFI. 
	For this energy resolution, we find $u=2.5$ to be the lowest interaction strength at which bipartite entanglement may be witnessed. 
	Panel (b) shows nQFI evaluated for $\eta\propto 1/L$, i.e. with size-dependent energy resolution. A suppression of nQFI is found for smaller systems, primarily at higher $u$ values. 
	Panel (c) shows the first derivative of the $L=128$ nQFI, calculated using both a standard forward finite difference (circles) and the Fornberg finite difference method \cite{Fornberg1988} (diamonds). Together these curves indicate a broad peak around $u=2.5$.
	}
\end{figure}
Having obtained the dynamical spin structure factors in the previous subsection, we now discuss the quantum Fisher information, a witness of multipartite entanglement. 
Figure~\ref{fig:QFI} shows the normalized quantum Fisher information (calculated from the spectra shown in Fig.~\ref{fig:skw}, as well as from spectra for smaller system sizes not shown) and its first derivative as functions of $u$. To avoid divergences as $T\rightarrow 0$, the QFI values were calculated using a small fictitious temperature of $k_B T = 0.001\tilde{t}$, below which the QFI was found to be approximately unchanged.

Fig.~\ref{fig:QFI}(a) shows results for $\eta=0.05\tilde{t}$. When the broadening $\eta$ is kept constant, all system sizes considered produce approximately equal nQFI values. This finding is consistent with the system size being larger than the witnessed entanglement depth. 
Under these conditions, we find that bipartite entanglement can be witnessed at $u\geq 2.5$. 

Under experimental conditions and at finite temperature, this cut-off likely moves to higher $u$ 
\footnote{At $u>5$ (i.e. beyond the plotted range) we find an unphysical flattening and even decay of nQFI, which is due to finite-size effects as discussed in Sec.~\ref{sec:results:DSF}. The expected physical behavior is that nQFI should approach the value obtained for the Heisenberg chain in the strong-coupling limit, which we find to be nQFI$\approx 2.2$, indicating at least tripartite entanglement in accord with prior results for the $S=1/2$ Heisenberg chain reported in Ref.~\cite{PhysRevB.103.224434}.}. 
Indeed, as has previously been discussed in the spin-system context, resolution limitations present a key challenge to experimental entanglement detection with QFI \cite{PhysRevB.103.224434, PhysRevLett.127.037201}. We investigate the resolution dependence for the Hubbard chain in Fig.~\ref{fig:QFI}(b) by leaving the broadening $\eta$ unchanged for $L=128$, and scaling it as $1/L$ for other system sizes. This results in a reduction of nQFI for smaller systems, which is negligible at the lowest $u$ values but becomes noticeable at intermediate $u$. The lowest $u$ at which bipartite entanglement is witnessed is $u=3.0$ for $L=64$, $u=2.75$ for $L=80$, $L=96$, and $u=2.5$ for $L\geq 112$. This nQFI reduction with $\eta$ is analogous to the finite-size effect described in the previous subsection, and its $u$-dependence is related to the $W(u)/\eta$ ratio. 

These results also suggest that nQFI can be increased by further improved resolution, which may shift the cut-off for observing bipartite entanglement to $u<2.5$. Limited frequency sampling and numerical errors cause uncertainties in the calculated QFI values (especially at lower $L$, higher $\eta$) that make finite-size scaling to the thermodynamic resolution limit unreliable. For this reason we leave determination of the theoretical cut-off value an open question.

Figure~\ref{fig:QFI}(c) shows the first derivative for the $L=128$ nQFI values. The derivative displays a broad peak around $u\approx 2.5$. The derivatives for smaller system sizes are consistent with the same broad peak, but are significantly ``noisier'' with  jumps from one $u$-value to the next, due to the aforementioned uncertainties.

\subsection{Van Hove correlations}
\begin{figure}
	\centering
	\includegraphics[width=\columnwidth]{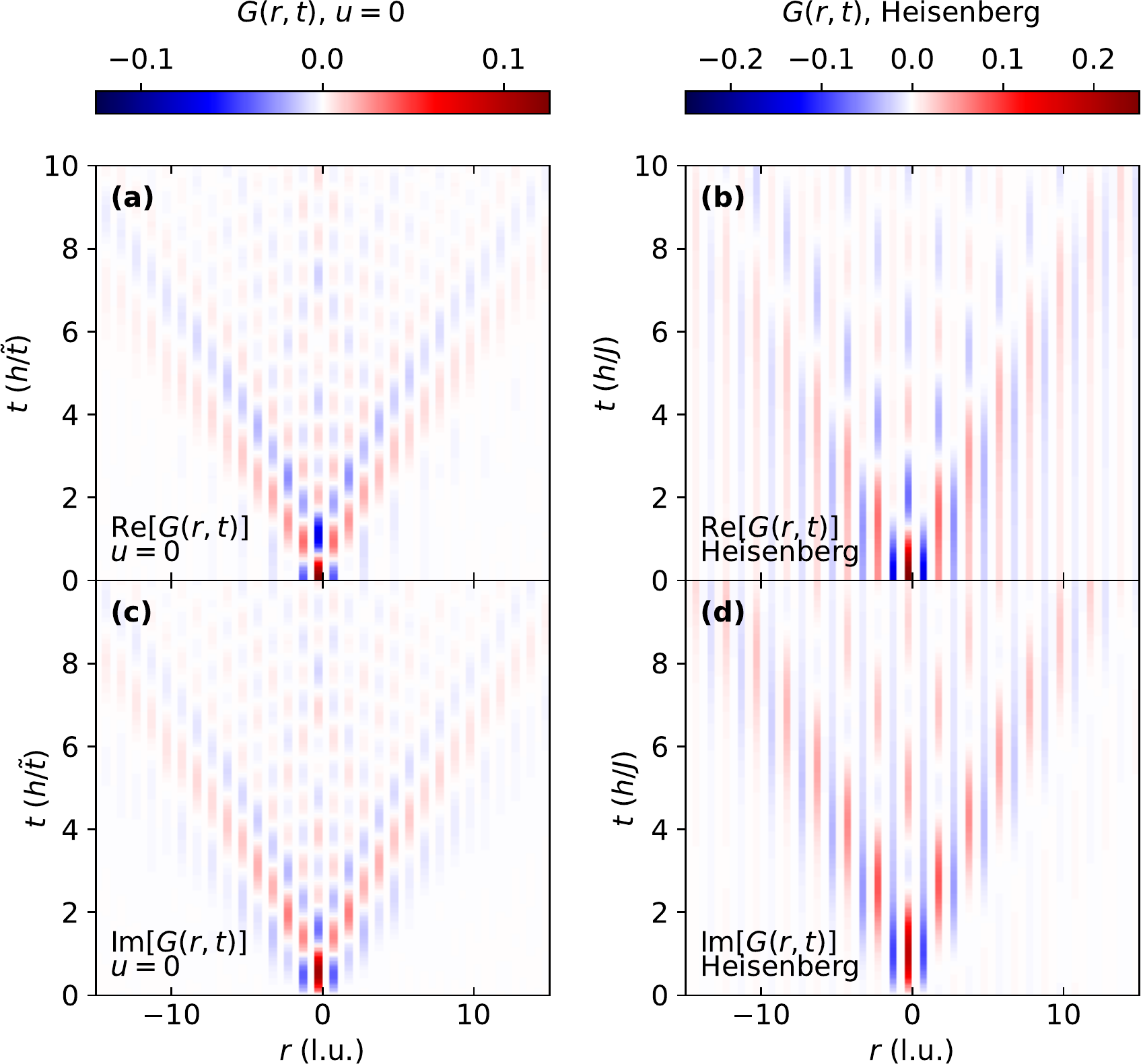}
	\caption{\label{fig:grt:twolimits}Real-time real-space correlations in the non-interacting (left column) and strong-coupling limits (right column). The top row panels show the real part, and the bottom row panels show the imaginary part. Ferromagnetic (antiferromagnetic) correlations $\langle S_r^z(t) S_0^z\rangle$ are indicated in red (blue).}
\end{figure}
\begin{figure}
	\centering
	\includegraphics[width=\columnwidth]{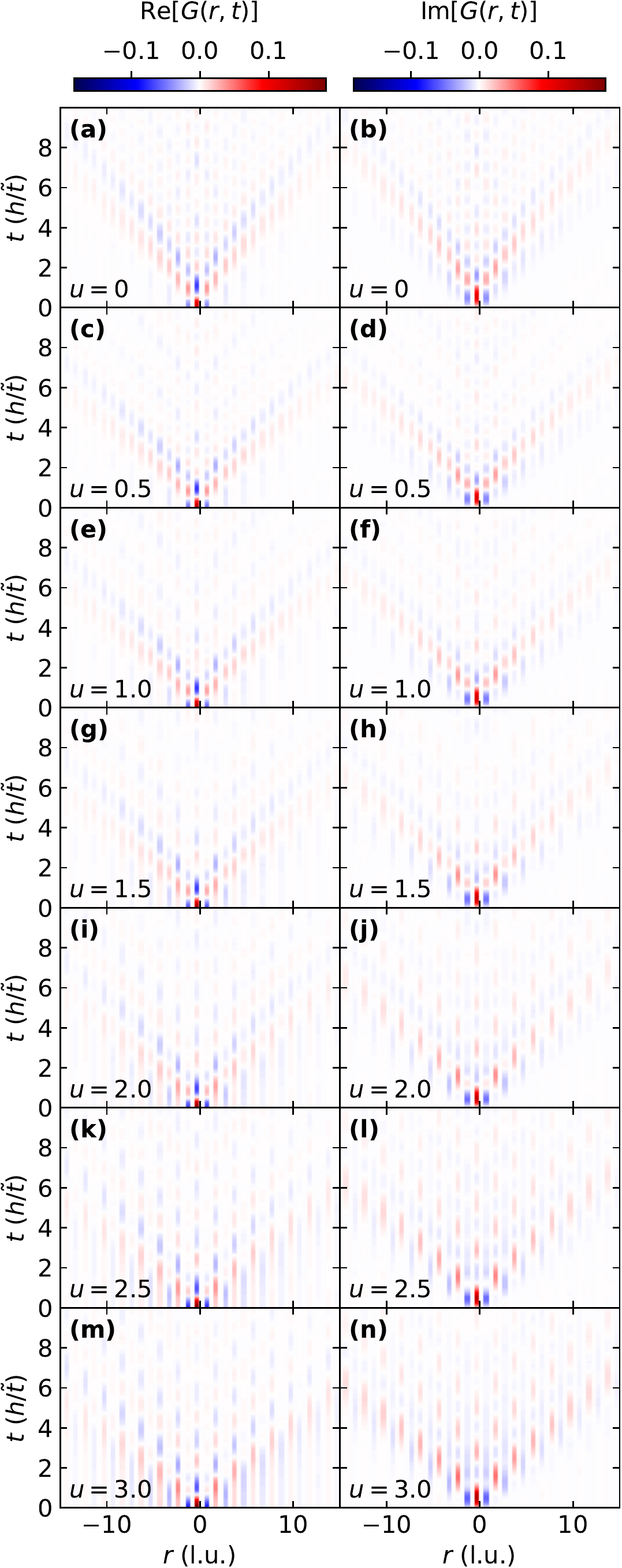}
	\caption{\label{fig:grt:lowU}Real-time real-space correlations at low to intermediate $u=U/\tilde{t}$. Left (right) column shows the real (imaginary part) of $\langle S_r^z(t) S_0^z\rangle$.}
\end{figure}
\begin{figure}
	\centering
	\includegraphics[width=\columnwidth]{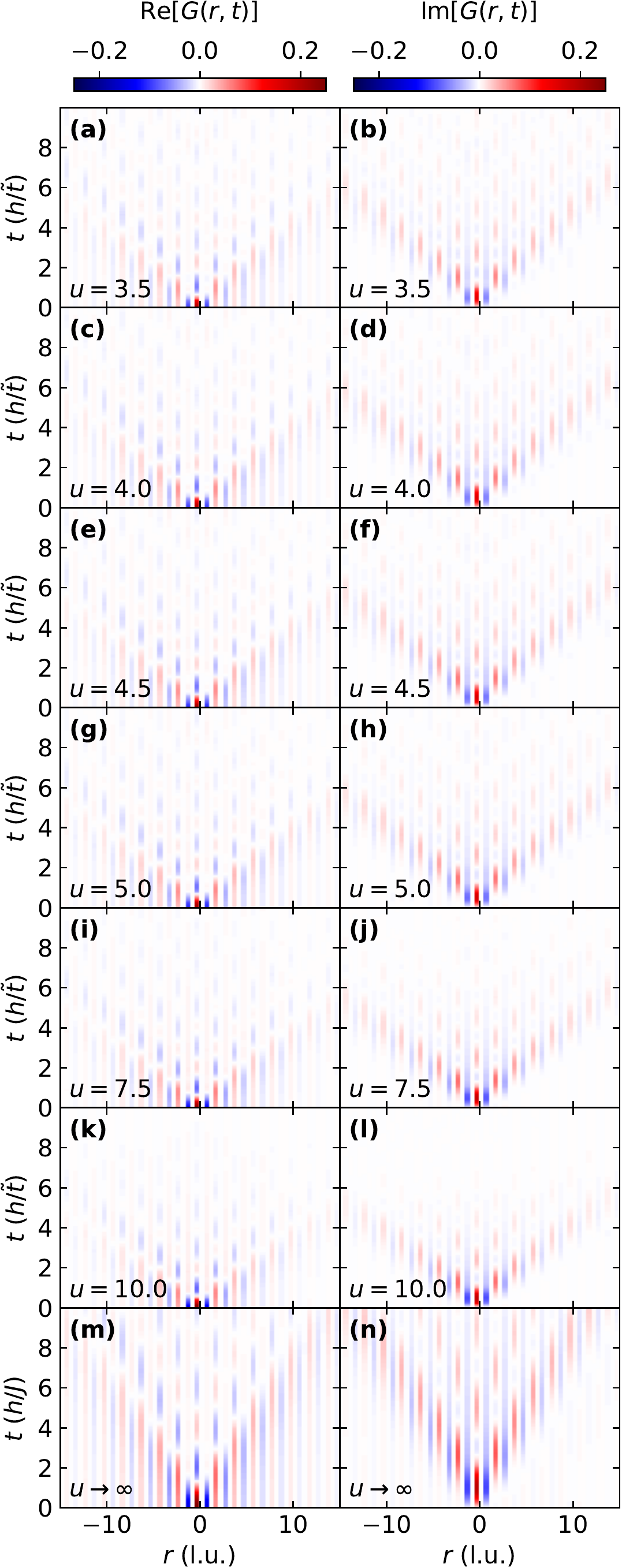}
	\caption{\label{fig:grt:highU}Real-time real-space correlations at intermediate to high $u=U/\tilde{t}$. Left (right) column shows the real (imaginary part) of $\langle S_r^z(t) S_0^z\rangle$.}
\end{figure}
In the previous subsections we have discussed properties in momentum and frequency space, $(k,\omega)$. We now turn to real-space real-time correlations. 
Figure~\ref{fig:grt:twolimits} shows the contrasting behavior of such Van Hove correlations in the non-interacting ($u=0$) and strong-coupling ($u\rightarrow\infty$) limits. The latter was previously studied in Ref.~\cite{Scheie2022}. In the strong-coupling limit there is a static background of $\mathrm{Re}[G(r,t)]$ due to non-zero N\'eel correlations in the groundstate. This background is absent at $u=0$. In both cases there is a ``light-cone'' controlling the propagation speed of correlations away from the $r=0$, $t=0$ origin, and timelike oscillations above it. In the Heisenberg case, the wavefront at the edge of the light-cone is characterized by AFM correlations. Above this cone, the system develops period-doubled $Q=\pi/2$ AFM correlations, while the nearest-neighbor correlations vanish. This feature was discussed in detail in Ref.~\cite{Scheie2022}, and explained in terms of interference of spinon quasiparticles. 
It was called a ``quantum wake'', in analogy with the smooth wake behind a moving ship. 
In the $u=0$ case we instead see ferromagnetic wavefronts, without any sign of period doubling. 

\begin{figure}
	\includegraphics[width=\columnwidth]{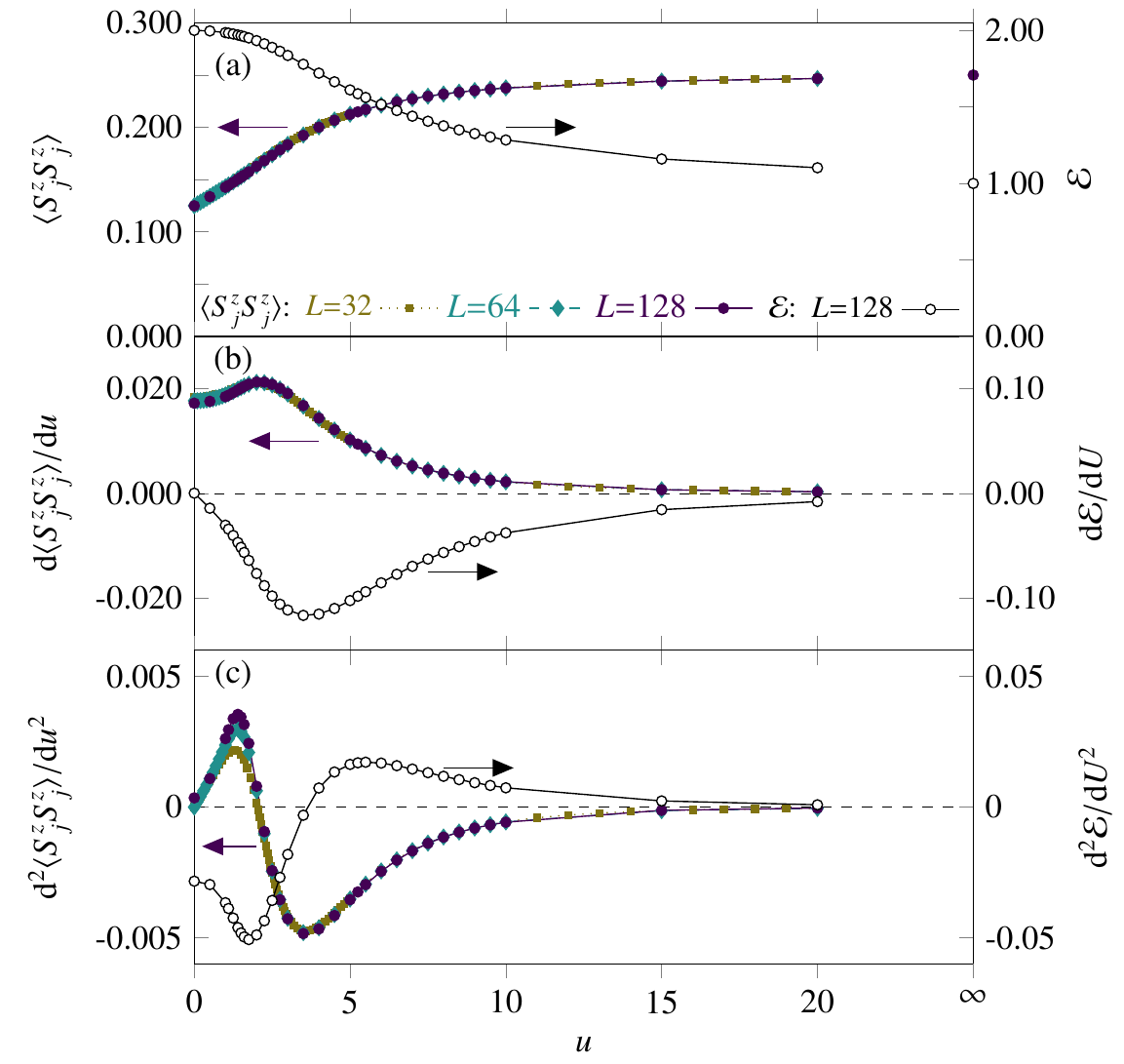}
	\caption{\label{fig:selfcorrnormalization}(a) Averaged on-site spin-spin correlation $\langle S_j^z S_j^z\rangle$ (filled symbols, left $y$-axis) and single-site entropy $\mathcal{E}$ (black line, open symbols, right $y$-axis). The single-site entropy indicates a reduction in the number of local basis states with $u$, which causes the on-site spin-spin correlation to grow as the weights of empty and doubly occupied states decrease. 
	(b) The first derivative of $\langle S_j^z S_j^z\rangle$ peaks near $u=2$ for all studied sizes, while the peak of $\mathrm{d}\mathcal{E}/\mathrm{d}u$ occurs at a higher value $u=3.5$, in agreement with Eq.~\eqref{singlesitenetropy:derivative}. 
	(c) The second derivative of $\langle S_j^z S_j^z\rangle$ changes sign near $u=2$ and has peaks near $u=1.4$ and $u=3.5$ for $L=64,128$ ($u=1.3$ and $u=3.6$ for $L=32$, which is sampled more densely). The second derivative of $\mathcal{E}$ peaks at $u=1.75$ and $u=5.5$, and changes sign near $u=3.5$. All derivatives were calculated using the Fornberg finite difference method \cite{Fornberg1988}.
	}
\end{figure}

The crossover between these two limits is evident in Figures~\ref{fig:grt:lowU},\ref{fig:grt:highU} showing results for $u\leq 3.0$ and $u>3.0$, respectively. The correlations are normalized such that $G(0,0)=\langle S^z_i S^z_i \rangle \leq 1/4$, where equality is reached at strong coupling, see Fig.~\ref{fig:selfcorrnormalization}(a). 
Already by $u\sim 2.5-3.0$ many of the features in $G(r,t)$ seen at strong coupling have developed, but not saturated. For example, Fig.~\ref{fig:grt:lowU}(m) shows nearly vanishing odd-neighbor correlations and timelike oscillations above the lightcone, a N\'eel-like static background, and an initially AFM wavefront at the lightcone. 
The crossover in $G(r,t)$ is smooth with $u$, and can be tracked through the strength of odd-neighbor correlations. Although the precise location of the crossover will depend on the choice of cutoff for these correlations, we note that at $u=2.0-2.5$ [panels (i),(k)] there remains visible FM odd-neighbor correlations above the lightcone. 
Thus, the dynamical correlations qualitatively approach the strong-coupling results for relatively modest values of $u$ also in the real-time, real-space domain.

\section{Discussion}\label{sec:discussion}
\subsection{Entanglement}
\begin{figure}
	\includegraphics[width=\columnwidth]{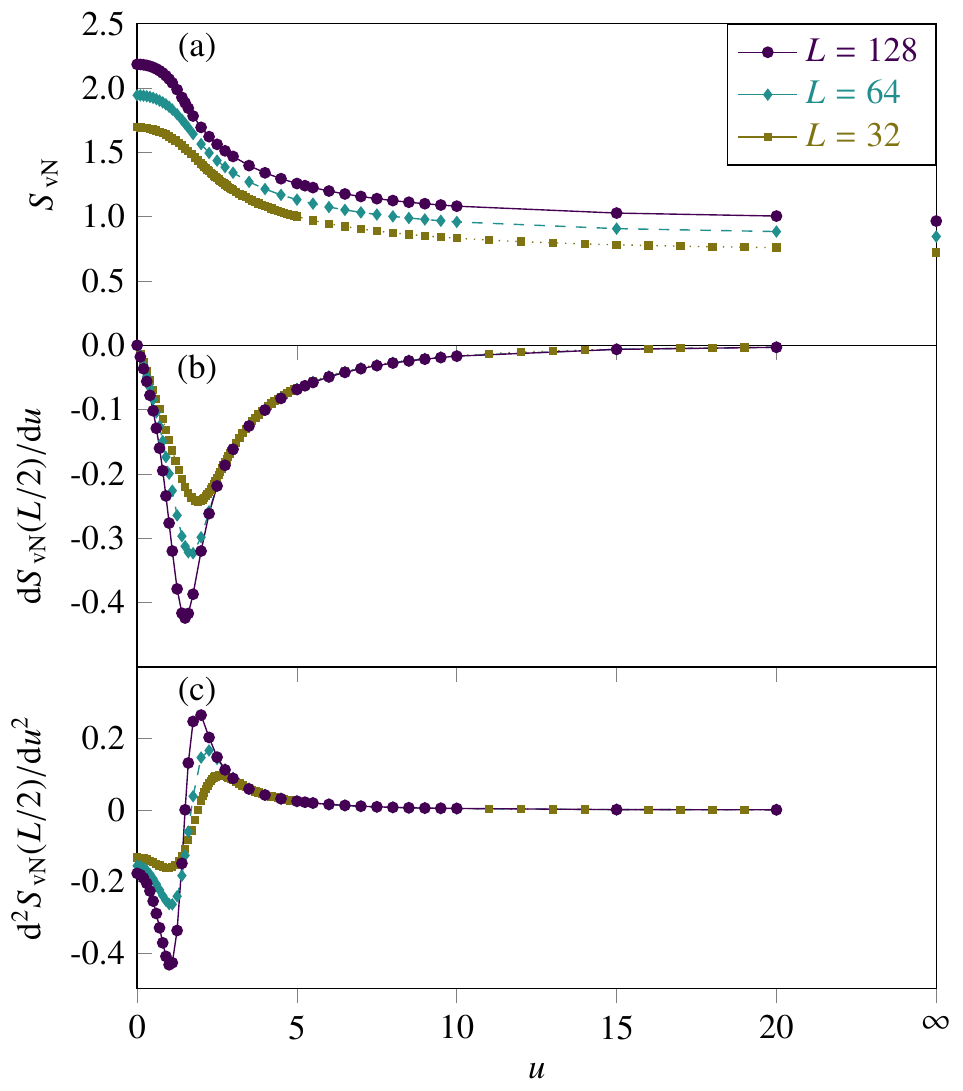}
	\caption{\label{fig:entanglemententropy}(a) The half-chain von Neumann entanglement entropy decreases as $u=U/\tilde{t}$ is increased, tending to the value for the Heisenberg chain as $u\rightarrow \infty$. 
	There is a change in curvature of the half-chain entropy at low $u$ below $u=2$. To see this we calculate the first and second derivatives in (b) and (c) using the Fornberg finite difference method \cite{Fornberg1988}.
	The first derivative peaks near $u=1.9$ for $L=32$, and near $1.5$ for $L=64$ and $L=128$.
	The second derivative changes sign near $u=1.8$ for $L=32$, $1.5$ for $L=64$, and $1.25$ for $L=128$. 
	The second peak in the second derivative occurs at $u=2.6$ for $L=32$, $2$ for $L=64$, and $1.75$ for $L=128$. 
	Although the shown data is very limited in scope, we find that naive finite-size scaling in $1/L$ is consistent with finite values for all three quantities. The finite-size scaling prediction is that the zero of the second derivative occurs at $u\approx 0.93$ and the second peak occurs at $u\approx 1.53$ in the thermodynamic limit. The peak in the first derivative is predicted to occur at $u=1.63$.
	}
\end{figure}

\begin{figure}
	\includegraphics[width=\columnwidth]{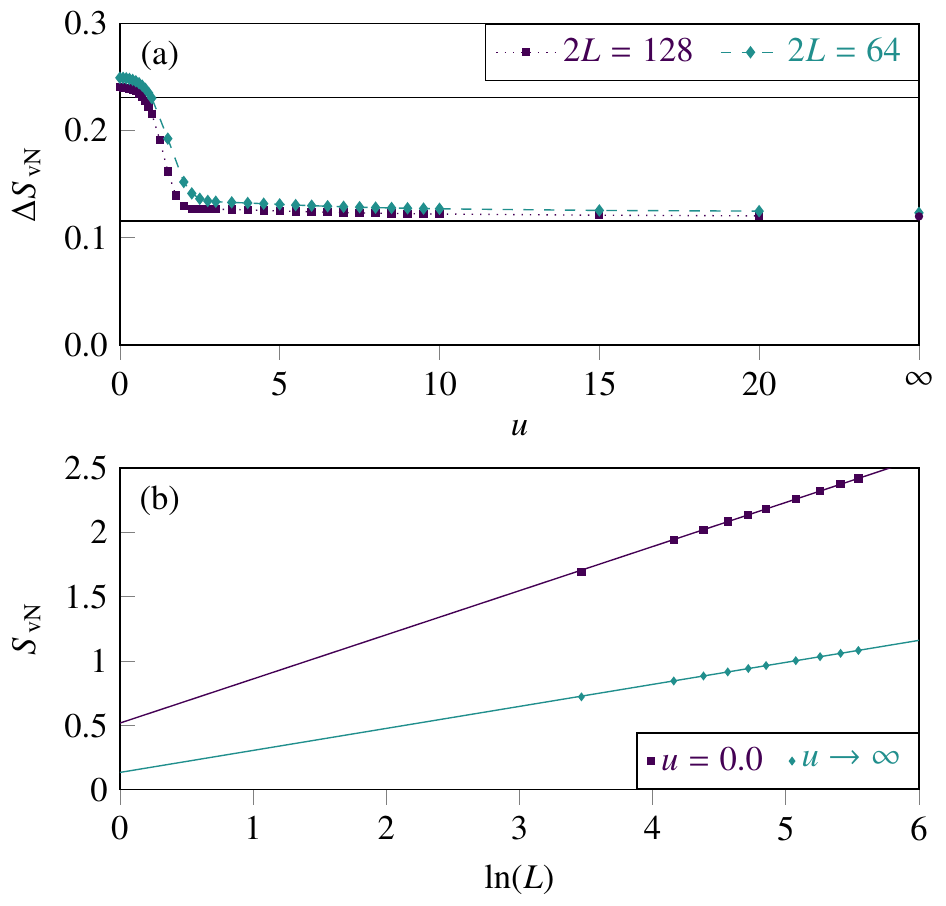}
	\caption{\label{fig:entanglemententropy:scaling}Reduction of central charge as function of $u$. (a) Plotted data series represent $\Delta S_\mathrm{vN}=S_\mathrm{vN}\left( 2L \right) - S_\mathrm{vN}\left( L \right)$. The horizontal lines are drawn at $\ln \left( 2\right)/3$ and $\ln \left( 2\right)/6$. 		
	(b) Finite-size scaling of $S_\mathrm{vN}$ at weak and strong coupling. The lines are linear least squares fits of the $L\geq 64$ entropies, with functional forms $0.518190 + 0.342946\ln(L)$ and $0.134746 + 0.171053\ln(L)$, respectively. The fits yield central charges $c^\mathrm{fit}_{u=0}\approx 2.058$ and $c^\mathrm{fit}_\mathrm{Heis}\approx 1.026$ and $c^\mathrm{fit}_\mathrm{Heis}/c^\mathrm{fit}_{u=0}\approx 2.005$, consistent with the expected $c_{u=0}=2$ and $c_\mathrm{Heis}=1$.
	}
\end{figure}

Our results show that QFI calculated from the dynamical spin structure factor can be experimentally used to witness at least bipartite entanglement for $u>2.5$ 
given sufficient energy resolution (assuming realistic values of $\tilde{t}\gtrsim 0.1$ eV, $\eta=0.05\tilde{t}\gtrsim 5$ meV is experimentally feasible). 
This cut-off may be shifted to somewhat lower $u$ with improved resolution. 
Yet we stress that the inability to witness entanglement at  
interaction 
strengths below the cut-off 
does not imply the absence of entanglement. Indeed, even non-interacting identical fermions are entangled because of the Pauli exclusion principle \cite{Vedral2003}. We have calculated the entanglement entropy, shown in Fig.~\ref{fig:entanglemententropy}, to demonstrate this explicitly. The entanglement entropy directly quantifies bipartite entanglement between two blocks of the system, and is found to decay with increasing $u$. The physical reason is simple \cite{PhysRevA.65.042101}: increasing $u$ implies suppression of charge fluctuations. Although the local Hilbert space has four states, eventually only two of them have appreciable weight. Increasing $u$ also results in increasingly prominent AFM spin correlations, which are then probed by our QFI formulation.

The reduction in the number of local basis states may be seen more directly through the lens of single-site entropy $\mathcal{E}$, which is defined in terms of the one-site reduced matrix. For the Hubbard chain it has been calculated analytically, yielding \cite{PhysRevLett.93.086402, PhysRevLett.95.196406, PhysRevA.74.042325}
\begin{align}
    \mathcal{E} &=  -w_0 \log_2 w_0 - w_\uparrow \log_2 w_\uparrow - w_\downarrow \log_2 w_\downarrow - D \log_2 D
\end{align}
where $D$ measures the double occupation and is given by Eq.~\eqref{eq:doubleoccupation}, and where
\begin{align}
    w_\sigma    &= \frac{1}{L} \sum_{i} \langle n_{i\sigma}\rangle-D, \quad \sigma=\uparrow,\downarrow,\\
    w_0 		&= 1 - w_\uparrow - w_\downarrow - D.
\end{align}
At half-filling and zero magnetization, $w_0=D$, and $w_\uparrow=w_\downarrow$ satisfies $\langle S_j^z S_j^z\rangle = w_\uparrow/2$, which leads to
\begin{align}
	\mathcal{E}&=	4 \langle S_j^z S_j^z\rangle \log_2 \left[ \frac{\frac{1}{2}-2\langle S_j^z S_j^z\rangle}{2\langle S_j^z S_j^z\rangle}\right] -\log_2 \left[ \frac{1}{2} -2\langle S_j^z S_j^z \rangle \right].	\label{eq:singlesiteentropy:SzSz}
\end{align}
The quantity $\mathcal{E}$ is thus plotted together with $\langle S_j^zS_j^z\rangle$ in Fig.~\ref{fig:selfcorrnormalization}(a). 
Similarly to the half-chain entropy in Fig.~\ref{fig:entanglemententropy}, the maximum of $\mathcal{E}$ is at $u=0$, which is followed by gradual decay as $u\rightarrow \infty$. 
As previously found by Refs.~\cite{PhysRevLett.93.086402, PhysRevA.74.042325} the finite-size dependence of $\mathcal{E}$ is negligible so only $L=128$ results are shown. 
The two-site entropy was calculated for low $u$ elsewhere \cite{PhysRevB.75.155108}, and also has a maximum at $u=0$. 

The numerical derivatives of $\mathcal{E}$ are plotted in Fig.~\ref{fig:selfcorrnormalization}(b),(c) along with derivatives of $\langle S_j^zS_j^z\rangle$. 
The first derivative of Eq.~\eqref{eq:singlesiteentropy:SzSz} is given by
\begin{align}
	\frac{\mathrm{d} \mathcal{E}}{\mathrm{d} u}	&=	4\log_2 \left[ \frac{1}{4\langle S_j^z S_j^z\rangle} - 1 \right] \frac{\mathrm{d} \langle S_j^z S_j^z \rangle }{\mathrm{d} u},	\label{singlesitenetropy:derivative}
\end{align}
which explains the behavior in Fig.~\ref{fig:selfcorrnormalization}(b). $\mathrm{d} \mathcal{E}/\mathrm{d} u$ vanishes at $u=0$ where $\langle S_j^zS_j^z\rangle=1/8$, and as $u\rightarrow \infty$ where $\mathrm{d} \langle S_j^z S_j^z\rangle/\mathrm{d} u =0$. At finite $u$ the $\log$ factor is negative and displaces the $\mathrm{d}\mathcal{E}/\mathrm{d}u$ peak to higher $u$ than the peak of $\mathrm{d}\langle S_j^zS_j^z\rangle/\mathrm{d}u$.

There is also a reduction in the number of gapless modes, from two decoupled spinless fermion species at $u=0$ to one spin degree of freedom at strong coupling. This results in a reduction in central charge from $c_{u=0}=2$ to $c_\mathrm{Heis}=1$. Our entanglement entropy results are consistent with this expectation. Conformal field theory predicts that the half-chain entanglement entropy in a system of length $L$ with open boundary conditions satisfies \cite{Calabrese2004}
\begin{align}
	S_\mathrm{vN}	&=	\frac{c}{6} \ln \left( \frac{L}{\pi}\right) + C,
\end{align}
where $C$ is a non-universal term that depends on correlation length and boundary corrections. If $C$ is independent of system size, the quantity
\begin{align}
	\Delta S_\mathrm{vN}	&=	S_\mathrm{vN} \left( 2L \right) - S_\mathrm{vN} \left( L \right)
\end{align}
reduces to $\frac{c}{6} \ln \left( 2\right)$. We plot $\Delta S_\mathrm{vN}$ in Fig.~\ref{fig:entanglemententropy:scaling}(a), finding that our $S_\mathrm{vN}$ approximately satisfy this relation at $u=0$ and as $u\rightarrow \infty$, with a gradual transition in-between. Minor deviations in the two limits are attributed to boundary corrections and truncation errors. Fig.~\ref{fig:entanglemententropy:scaling}(b) shows that the entropies at $u=0$ and strong coupling scale logarithmically with $L$. Central charges obtained from $S_\mathrm{vN}$ fits are consistent with the theoretical values.

We also find that derivatives of the entanglement entropy [Fig. \ref{fig:entanglemententropy}(b),(c)] show a crossover at low $u$, similarly to QFI and Van Hove correlations, albeit at weaker interactions. This likely indicates the rapid suppression of charge fluctuations, whereas the growth of QFI also depends on the build-up of spin-spin correlations. We note that QFI can be defined for arbitrary bounded and Hermitian operators --- the choice of spin operators is to allow predictions for neutron scattering. A different choice of operators may be able to witness a higher degree of multipartite entanglement. That is indeed seen in a recent study \cite{PhysRevResearch.3.L032051} introducing a quench protocol for measuring QFI, which is more suitable to ultracold fermionic gases and quantum Hall devices than general condensed matter systems. In general, it would be valuable to have additional experimentally accessible entanglement measures for correlated electron systems.

\subsection{Crossover}
We find that $S(k,\omega)$, QFI, $G(r,t)$ and entanglement entropy all display crossovers at $u<W(0)$. $\mathrm{Re}[G(r,t)]$ directly indicates the build-up of short-range N\'eel correlations, which result in $S(k,\omega)$ qualitatively approaching the spectrum for the Heisenberg chain. At the same time, the bandwidth of $S(k,\omega)$ contracts in a manner that shows a further crossover near $u=2.25$, see Fig.~\ref{fig:bandwidth:bethe} and Appendix~\ref{app:bandwidth}. The combined bandwidth contraction and build-up of AFM correlations (reflected by the $S(k=\pi,\omega)$ peak) results in the QFI crossover near $u=2.5$. These values are consistent with the previously made observation \cite{PhysRevB.75.205128, PhysRevB.94.205145} that $S(k,\omega)$ has qualitatively approached its strong-coupling form by $u=3$.

Although the locations of the crossover points are found to vary between these quantities, they all reflect an underlying trend from itinerant to localized behavior as $u$ is increased. The decay of entanglement entropy with $u$ provides clear support for this picture. Our results also lend support to an experimental rule of thumb that systems may be considered more electronic for $u \lesssim 2$ and magnetic for $u \gtrsim 2$. 
These results suggest that the transition from weak to strong coupling regimes occurs at values of $u$ smaller than naively anticipated considering that the bandwidth of the noninteracting model is $W(0)=4\tilde{t}$.

\subsection{Experimental considerations}
In a general sense, our results demonstrate that the QFI and $G(r,t)$ analysis of Refs.~\cite{PhysRevB.103.224434, PhysRevLett.127.037201, Scheie2022} may be applied to systems with electronic degrees of freedom. 
A quantitative determination of QFI may require theoretical modeling in order to ensure correct normalization, however $G(r,t)$ need not be normalized to yield useful insights about the local dynamics. Although we have assumed spin SU(2) symmetry throughout this work, spin anisotropy should be possible to handle using the approaches laid out in Ref.~\cite{PhysRevLett.127.037201}.

Directly observing crossovers in correlations by tuning $u$ is not possible in materials, but may be feasible using quantum simulator platforms \cite{doi:10.1126/science.aag1635, Tarruell2018}. The situation in real quasi-one-dimensional materials is typically also complicated by the presence of additional orbitals, interactions and hopping paths \cite{Refolio_2005,PhysRevB.56.3402}, which influence the physics, particularly at low temperature. 
Nevertheless, it is well-established that the high-energy inelastic neutron scattering can sometimes be quantitatively described by simplified models. This occurs, for example, in large-$u$ systems such as KCuF$_3$ \cite{PhysRevB.103.224434, PhysRevLett.111.137205} and SrCuO$_2$ \cite{PhysRevLett.93.087202}, for which the magnetic excitations are well-captured by the Heisenberg model. In principle, it may be possible to identify systems exemplifying various values of $u$. 

Perusal of the literature reveals a lack of clearcut examples of material realizations of low-$u$ Hubbard chains suitable to study with neutrons. The Mott-insulating organic Bechgaard-Fabre salts \cite{Dressel2003} are thought to be away from strong coupling, but also have low neutron cross sections due to the dilute nature of the spin moments in the materials which are further obscured by the large amount of scattering from hydrogen. A recently proposed solid state candidate is Ti$_4$MnBi$_2$ \cite{PhysRevB.102.014406}, for which inelastic data currently does not exist. 
Given this lack of clearcut options, it may be more promising to look for weakly coupled ladder systems. It is presently unclear to which extent our findings for the Hubbard chain will translate to such ladders, but we note that Ref.~\cite{PhysRevB.94.205145} previously found that both half-filled Hubbard chains and ladders show crossovers of $S(k,\omega)$ at low $u$. If our findings can be generalized to doped systems, more possibilities open up. Some examples of such ladder compounds include (TaSe$_4$)$_2$I with estimated $u\sim 1$ \cite{Refolio_2005, PhysRevB.101.174106} and K$_{0.3}$MoO$_3$ (molybdenum blue-bronze) with $u\sim 4$ \cite{PhysRevB.89.201116, Refolio_2005}, both of which develop a charge density order at low temperatures, and Li$_{0.9}$Mo$_6$O$_{17}$ (lithium purple bronze) \cite{PhysRevB.68.195117, Dudy_2012, PhysRevB.89.045125}, which was estimated to be in the weak-coupling regime but potentially is better understood in a multiorbital model \cite{PhysRevB.92.134514}.

An intriguing finding here is that the van Hove correlations show the development of local versus itinerant magnetism as a consequence of correlations. Additionally, the QFI indicates the need for a quantum description. This suggests that van Hove correlations and QFI garnered from neutron scattering experiments, provided that $S(\mathbf{k},\omega)$ is measured to sufficient accuracy to extract these quantities, could provide a useful viewpoint for interpreting the states in correlated itinerant materials more generally, including in higher dimensions. 
Examples where insight may be gained are unconventional superconductors where magnetism plays an important competing role. Indeed suitable data in terms of wave-vector and energy coverage may already be available and it would be interesting to compare trends within materials classes with doping and composition. 

Finally, the charge fluctuations themselves can be anticipated to also contain insightful information. While the non-Hermitian nature of the creation and annihilation operators in $A(\mathbf{k},\omega)$ 
is 
problematic, dynamical charge density correlations $N(\mathbf{k},\omega)$ could be of interest and experimentally accessible through Bragg scattering \cite{Ernst2010} in cold atom systems or electron energy loss spectroscopy (EELS) in solid-state systems \cite{SciPostPhys.3.4.026}. A study to explore this is planned and the prospects for experimental measurements are to be considered.

\section{Conclusion}\label{sec:conclusion}
We have studied the magnetic excitations of the half-filled Hubbard chain from low to intermediate $u$, as measured in units of the noninteracting bandwidth. We find that the dynamical spin structure factor, quantum Fisher information and Van Hove correlations all display crossovers at low $u$ that are attributed to the more fundamental crossover from itinerant electron physics to localized spin physics. This is reflected directly in the build-up of N\'eel correlations, as may be seen in the Van Hove correlations. This suggests that Van Hove correlation analysis of neutron scattering data is of interest also in charged systems. In addition, we have shown how to adapt QFI derived from $S(k,\omega)$ to models with electronic degrees of freedom, finding that bipartite entanglement may be witnessed above $u\approx 2.5$ with realistic energy resolution. Our results thus present one path to entanglement quantification in correlated electron system that is applicable beyond the quantum \emph{spin} systems previously studied \cite{PhysRevB.103.224434, PhysRevLett.127.037201}.

\begin{acknowledgments}
	We thank A. Nocera for helpful discussions about the Krylov correction vector algorithm. 
	The work of PL, SO, and ED was supported by the U.S. Department of Energy (DOE), Office of Science, Basic Energy Sciences (BES), Materials Sciences and Engineering Division. AS was supported by the DOE Office of Science, Basic Energy Sciences, Scientific User Facilities Division. 
	The work by DAT is supported by the Quantum Science Center (QSC), a National Quantum Information Science Research Center of the U.S. Department of Energy (DOE). GA was supported in part by the scientific Discovery through Advanced Computing (SciDAC) program funded by U.S. Department of Energy, Office of Science, Advanced Scientific Computing Research and Basic Energy Sciences, Division of Materials Sciences and Engineering, and in part by the ExaTN ORNL LDRD. Software development has been partially supported by the Center for Nanophase Materials Sciences, which is a DOE Office of Science User Facility.
\end{acknowledgments}

\appendix
\section{Some properties of \texorpdfstring{$G(r,t)$}{G(r,t)}}\label{app:properties}
In this appendix we collect additional properties of the Van Hove correlation functions.

In general, $G(r,t)	=	G^\star (-r,-t)$ since $S(k,\omega)$ is real-valued. For inversion symmetric systems with $G(r,t) =	G(-r,t)$ it follows that $\mathrm{Re}[G(r,t)]$ is even in $t$ and $\mathrm{Im}[G(r,t)]$ is odd in $t$. 

Note that detailed balance would be broken if $G(r,t)$ were real and even in $t$. Since $S(k,\omega)\sim \int G(r,t) e^{-ikr}e^{-i\omega t}\mathrm{d}r\mathrm{d}t$, this would imply $S(k,\omega)=S(-k,-\omega)$, which only holds as $T\rightarrow \infty$. 
Detailed balance may also be used to derive the relation 
\cite{PhysRevLett.4.239, Egelstaff1965}
\begin{align}
	\mathrm{Im}[ G(r,t)] 	&= - \tan \left(\frac{\hbar}{2k_B T} \frac{\partial}{\partial t}\right) \mathrm{Re}[G(r,t)].
\end{align}

By writing 
\begin{align}
G(r,t)	&=	\mathrm{Re}[G(r,t)] + i\mathrm{Im}[G(r,t)] = a(r,t) + ib(r,t),
\end{align}
we may obtain the Fourier transforms of the individual functions $a(r,t)$, $b(r,t)$
\begin{align}
	a(k,\omega)	&=	\frac{S(k,\omega) + {S^{\star}(-k,-\omega)}}{2},	\\
	b(k,\omega)	&=	\frac{S(k,\omega) - {S^{\star}(-k,-\omega)}}{2i}.
\end{align}
Using that $S(k,\omega)$ is real-valued and even in $k$, and detailed balance, we obtain 
the ratio
\begin{align}
\frac{S(k,\omega)}{a(k,\omega)}	&=	2\left( 1- \frac{e^{-\beta \omega}}{1+e^{-\beta \omega}}\right)
\end{align}
which approaches $1$ as $T\rightarrow \infty$, and $2$ as $T\rightarrow 0$, and
\begin{align}
\frac{S(k,\omega)}{b(k,\omega)}	&=	 2i\left( 1+ \frac{e^{-\beta \omega}}{1-e^{-\beta \omega}}\right)
\end{align}
which diverges as $T\rightarrow \infty$ and approaches $2i$ as $T\rightarrow 0$.

\section{Bandwidth renormalization}\label{app:bandwidth}
\begin{figure}
	\includegraphics[width=\columnwidth]{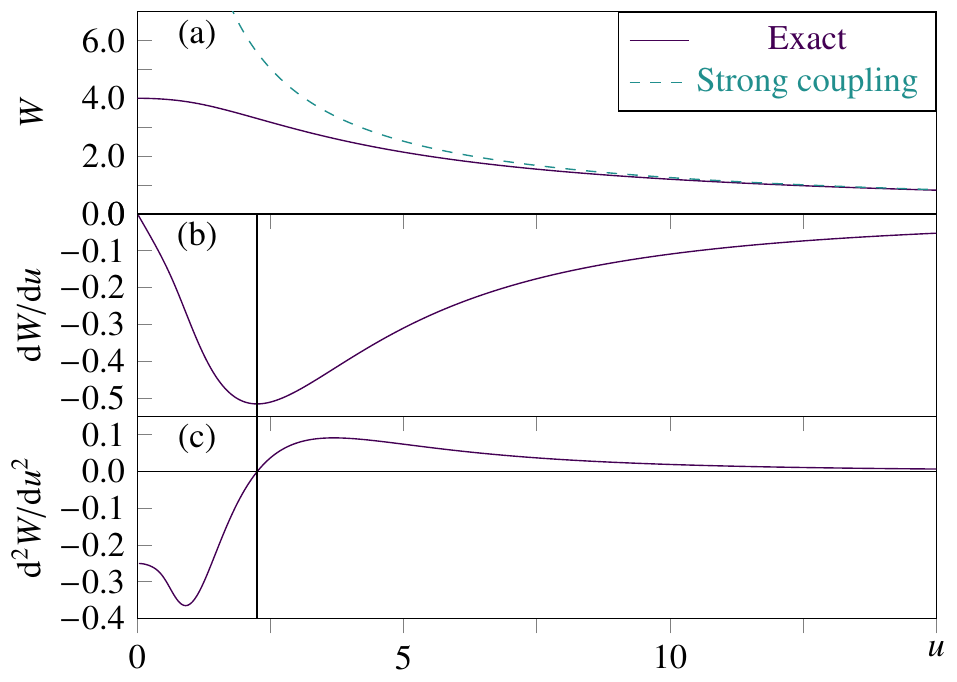}
	\caption{\label{fig:bandwidth:bethe}
		(a) Bandwidth $W(u)$ with $u=U/\tilde{t}$ according to the strong-coupling prediction ($u\rightarrow \infty$) and the Bethe ansatz. The difference becomes asymptotically small as $u\rightarrow \infty$. (b) The first derivative of $W(u)$ peaks near $u=2.25$, while (c) the second derivative changes sign near $u=2.25$. This value is marked by the vertical line in panels (b)-(c).
	}
\end{figure}
\begin{table}[tb]
	\caption{\label{table:hubbardbandwidth}Bandwidth, predicted and effective Heisenberg couplings, and scale factor as a function of $u=U/\tilde{t}$. The column labeled $W(u)$ shows the bandwidth in the exact solution, while $\pi J$ indicates the bandwidth predicted by the lowest order strong-coupling expression. The column labeled $J$ shows the Heisenberg coupling predicted to lowest order at strong coupling, $J=4\tilde{t}^2/U$. $J_\mathrm{eff}=W(u)/\pi$ is an effective Heisenberg coupling, chosen to reproduce the exact bandwidth. The column labeled ``Scale factor'' provides a number by which all energy scales can be scaled in order to match the non-interacting bandwidth.}
	\begin{tabular}{l|ll|ll|l}\hline
		$u$ & $W(u)$ & $\pi J$ & $J$	& $J_\mathrm{eff}$ & Scale factor \\\hline
		$0$		& $4$			& ---					& ---			&	$1.2732395$			& $1$		\\
		$0.5$		& $3.9680598$			& $25.132741$					&$8$		&	$1.2630727$			& $1.00805$		\\
		$1$ 	& $3.8615201$   &   $12.566371$ 	& $4$       & $1.22916001$ 	& $1.03586$               \\
		$1.5$		& $3.6709075$			& $8.3775804$					&	$2.67$		&$1.1684862$				& $1.08965$		\\
		$2$ 	& $3.4280008$   &   $6.2831853$     & $2$       & $1.09116655$ 	& $1.16686$           \\
		$2.25$	&	$3.2993560$	&	$5.5850536$		&	$1.778$	& $1.05021764$	&	$1.21236$	\\
		$2.5$		& $3.1706954$			&	$5.0265482$				&	$1.6$		&	$1.0092637$			& $1.26155$		\\
		$3$		& $2.9218858$	&	$4.1887902$ 	& $1.33$	& $0.93006515$ 	& $1.36898$ \\
		$3.5$		& $2.6920393$			&	$3.5903916$				&	$1.14$		&	$0.85690273$			& $1.48586$		\\
		$4$		& $2.484563$			&	$3.1415927$				&	$1$		&	$0.79086096$			& $1.60994$		\\
		$4.5$		& $2.2993673$			&	$2.7925268$				&	$0.889$		&	$0.73191136$			& $1.73961$		\\
		$5$		& $2.1348205$	&	$2.5132741$		& $0.8$		& $0.67953446$ 	& $1.87369$ \\
		$7.5$	& $1.5468125$	&	$1.6755161$		& $0.533$	& $0.49236571$	& $2.58596$	\\
		$10$	& $1.1992643$	&	$1.2566371$ 	& $0.4$		& $0.38173768$ 	& $3.33538$ \\
		$15$	& $0.8200271$	&	$0.8377580$		& $0.267$	& $0.26102272$ 	& $4.87789$  \\
		$20$	& $0.6207225$	&	$0.6283185$ 	& $0.2$		& $0.19758209$ 	& $6.44410$	\\
		$30$	& $0.41660302$	&	$0.41887902$ 	& $0.133$		& $0.13260886$ 	& $9.60147$ \\\hline
	\end{tabular}
\end{table}
The bandwidth $W(u)$ shrinks as $u$ is increased. At large $u$ it tends towards the upper limit of the two-spinon continuum of the Heisenberg chain, $\pi J \sin(k/2)\sim \pi 4\tilde{t}^2 \sin(k/2)/U$. At low $u$, however, the bandwidth is smaller than predicted by this strong-coupling expression, due to softening of spin excitations by moving charges \cite{Choy_1982}. A corrected spectrum of $S=1$ ``spin-wave'' excitations can be obtained using Bethe ansatz methods \cite{Choy_1982, Kluemper1990}. At half-filling, Ref.~\cite{Choy_1982} derived the expressions
\begin{align}
	\epsilon\left(\alpha,\beta\right)	&=	\frac{4\tilde{t}}{u} \int_{-\pi/2}^{\pi/2} \mathrm{d}k \cos^2(k) \left\{ \mathrm{sech}\left[ \frac{2\pi}{u}\left( \sin(k) - \alpha\right)\right] \right.\nonumber\\
							&\left. + \mathrm{sech}\left[ \frac{2\pi}{u}\left( \sin(k) - \beta\right)\right] \right\},
\end{align}
\begin{align}
	P\left(\alpha,\beta\right)						&=	\frac{2}{\pi}	\int_{-\pi/2}^{\pi/2} \mathrm{d}k \left\{ \tan^{-1} \left[ \exp \left( -\frac{2\pi}{u} \left( \alpha - \sin(k)\right)\right)\right] \right.\nonumber\\
							&\left. +  \tan^{-1} \left[ \exp \left( -\frac{2\pi}{u} \left( \alpha - \sin(k)\right)\right)\right] \right\}
\end{align}
for allowed energies and momenta, respectively. The full spectrum is obtained by varying the real numbers $\alpha,\beta$, which may be considered `holes' in the so-called $\Lambda$ distribution internal to the Bethe ansatz solution \cite{PhysRevLett.20.1445}. The $k$s represent pseudomomenta of said `holes'. 
The energy satisfies $\epsilon(\alpha,\beta)=\epsilon(-\alpha,-\beta)$, and reaches its minimum as $\alpha\rightarrow \infty$ and $\beta\rightarrow \infty$, for which $\epsilon(\alpha,\beta)\rightarrow 0$ and $P(\alpha,\beta)\rightarrow 0$. The energy maximum is reached at $\alpha=\beta=0$, 
\begin{align}
	\epsilon_\mathrm{max} &=	\epsilon(0,0) = \frac{8\tilde{t}}{u} \int_{-\pi/2}^{\pi/2} \mathrm{d}k \cos^2(k) \mathrm{sech}\left[ \frac{2\pi}{u} \sin(k) \right], \label{eq:bandwidth}
\end{align}
which corresponds to $P(0,0)=\pi$, i.e. the AFM wave vector. Thus the bandwidth $W(u)=\epsilon_\mathrm{max}$, which may be evaluated numerically. At strong coupling ($u\rightarrow \infty$), the integral evaluates to $\pi/2$, recovering the strong-coupling bandwidth result $4\pi |\tilde{t}|/u = 4\pi \tilde{t}^2/U= \pi J$. The bandwidth along with its derivatives is plotted in Fig.~\ref{fig:bandwidth:bethe}. 

Numerical values of the bandwidth predicted by Eq.~\eqref{eq:bandwidth} and the second-order strong-coupling expansion are shown in Table~\ref{table:hubbardbandwidth}. Also shown is the strong-coupling value of the Heisenberg exchange, $J=4\tilde{t}^2/U$, and an effective Heisenberg coupling, $J_\mathrm{eff}$, that reproduces the bandwidth found by the Bethe ansatz. The deviation between the two values $J$ and $J_\mathrm{eff}$ is significant at low $u$. At large enough $u$, this $J_\mathrm{eff}$ may be used for fitting experimental dispersions in a M\"uller ansatz \cite{Muller1981} approach, as suggested in Ref.~\cite{PhysRevB.71.020405}. However, such an approach would only be approximate as itineracy effects modify the high-energy scattering. Below $u\lesssim 3$, where the spectrum is qualitatively different from the strong-coupling limit, the M\"uller ansatz is insufficient. Table~\ref{table:hubbardbandwidth} also shows the scale factor used for all energy scales ($u$, $\eta/\tilde{t}$) to achieve constant bandwidth in Fig.~\ref{fig:skw:scaled}.


%

\end{document}